\documentclass[conference]{IEEEtran}
\usepackage{cite}
\usepackage{amsmath,amssymb,amsfonts}
\usepackage{algorithmic}
\usepackage{graphicx}
\usepackage{textcomp}
\usepackage{xcolor}

\usepackage{csquotes}
\usepackage{multirow}
\usepackage{hyperref}
\usepackage{cleveref}   
\usepackage{makecell}
\usepackage{fixltx2e}
\usepackage{breakurl}
\usepackage{bm}

\begin{document}

\def\UrlBreaks{\do\/\do-}
\newcommand{\comment}[1]{}

\title{Newsalyze: Effective Communication of Person-Targeting Biases in News Articles}

\author{\IEEEauthorblockN{Felix Hamborg\IEEEauthorrefmark{1}\IEEEauthorrefmark{3},
Kim Heinser\IEEEauthorrefmark{3}, Anastasia Zhukova\IEEEauthorrefmark{4}, 
Karsten Donnay\IEEEauthorrefmark{1}\IEEEauthorrefmark{2} and Bela Gipp\IEEEauthorrefmark{1}\IEEEauthorrefmark{4}}
\IEEEauthorblockA{\IEEEauthorrefmark{1}Heidelberg Academy of Sciences and Humanities, Germany,  \IEEEauthorrefmark{3}University of Konstanz, Germany, \\}
\IEEEauthorblockA{\IEEEauthorrefmark{2}University of Zurich, Switzerland, \IEEEauthorrefmark{4}University of Wuppertal, Germany}
felix.hamborg@uni-konstanz.de
}

\maketitle

\begin{abstract}
Media bias and its extreme form, fake news, can decisively affect public opinion. Especially when reporting on policy issues, slanted news coverage may strongly influence societal decisions, e.g., in democratic elections. Our paper makes three contributions to address this issue. First, we present a system for bias identification, which combines state-of-the-art methods from natural language understanding. Second, we devise bias-sensitive visualizations to communicate bias in news articles to non-expert news consumers. Third, our main contribution is a large-scale user study that measures bias-awareness in a setting that approximates daily news consumption, e.g., we present respondents with a news overview and individual articles. We not only measure the visualizations' effect on respondents' bias-awareness, but we can also pinpoint the effects on individual components of the visualizations by employing a conjoint design. Our bias-sensitive overviews strongly and significantly increase bias-awareness in respondents. Our study further suggests that our content-driven identification method detects groups of similarly slanted news articles due to substantial biases present in individual news articles. In contrast, the reviewed prior work rather only facilitates the visibility of biases, e.g., by distinguishing left- and right-wing outlets. 
\end{abstract}


\section{Introduction}
\label{sec1}
News articles serve as an essential source of information on current events. While a rich diversity of opinions is, of course, desirable also in the news, systematically biased information can be problematic as a basis for collective decision-making, e.g., in democratic elections, if not recognized as such. Empowering newsreaders in recognizing biases in coverage, especially on policy issues, is thus crucial. 

We take this as a motivation to define our research question: \textit{How can we effectively communicate instances of bias in a set of news articles reporting on the same political event to non-expert news consumers?} The main contributions of this paper are the first \textit{large-scale user study (contribution C1)} that employs a conjoint design to investigate how visualizations and also individual components therein can help news consumers to become aware of biases. Our study employs a setting that resembles daily news consumption. Before the study, we introduce our \textit{bias identification system (C2)}, which combines state-of-the-art methods to identify media bias. Moreover, we introduce layouts and components to build \textit{modular visualizations to communicate biases (C3)}.

We publish the survey materials, including questionnaires, articles, and visualizations: \url{https://zenodo.org/record/4704891}

\section{Related Work}
\label{sec2}

\subsection{Definitions}
\label{sec2:definitions}
Defining media bias is a challenging task \cite{Park2009b} due to overlapping or even contrary bias theories \cite{Hamborg2019d}. We rely on a frequent (though not consistent) concept among bias definitions developed by social science researchers, where \textit{media bias} is a relative concept, i.e., bias can only be evaluated comparatively to other information, e.g., news articles (cf. \cite{Elejalde2018,Pitoura2018,Hamborg2019d}). 

We define \textit{bias-awareness} generally as an effect of bias communication. In practical terms, we define bias-awareness in this paper as an individual's motivation and ability to relate and contrast perspectives present in news coverage to another \cite{Park2009b} and also to the individual's views \cite{Giner-Sorolla1994}.

\subsection{Approaches}
\label{sec2:approaches}
We limit our problem statement to the analysis and communication of media bias (also called bias diagnosis, measurement, and mitigation \cite{Park2009b}). We exclude other means to address media bias, such as bias prevention during news production \cite{Park2009b}. Bias-sensitive visualizations may support news consumers in making more informed choices \cite{Baumer2017}. However, we find that the prior work suffers from the following shortcomings. 

\textit{High cost and lack of recency:} Content analyses and frame analyses are among the most effective bias analysis tools. Decade-long research in the social sciences has proven them effective and reliable, e.g., to capture also subtle yet powerful biases (cf. \cite{entman1993framing,Hamborg2019d}). However, because researchers need to conduct these analyses mostly manually, the analyses do not scale with the vast amount of news \cite{Hamborg2019d}. In turn, such studies are always conducted for (few) topics in the past and do not deliver insights for the current day \cite{McCarthy2008,Oliver2000}; this would, however, be of primary interest to people reading the news.

\textit{Superficial results:} Many automated approaches suffer from superficial results, especially when compared to the results of cumbersome analyses as conducted in the social sciences \cite{Hamborg2019d}. Reasons include that the approaches treat media bias as a rather vaguely or broadly defined concept, e.g., ``differences of [news] coverage'' \cite{park2011politics},
``diverse opinions'' \cite{munson2010presenting}, or ``topic diversity'' \cite{Munson2009}. Further, especially early approaches \cite{Park2009b,Munson2009} suffer from poor performance since word-, dictionary-, or rule-based methods as commonly employed in traditional machine learning fail to capture the ``meaning between the lines'' \cite{Hamborg2019d}. To improve performance, some approaches employ crowdsourcing \cite{Park2011,AllSides.com2021,Spinde2020}, e.g., to gain bias ratings. Crowdsourced data can be an effective means to gather labeled data. However, such data can be problematic if not carefully reviewed for biases \cite{Hube2019}, e.g., if users are not a representative sample or already biased through earlier exposure to systematically biased news coverage. Recent methods that employ deep learning or word embeddings can yield more substantial results. However, the creation of large-scale datasets required for their training is very costly \cite{Card2015,Hamborg2021b}, and semi-automated approaches require careful, manual revision of the automatically identified bias categories \cite{Kwak2020}.

\textit{Inconsistency:} The design of some approaches only facilitates the visibility of biases that might be in the data rather than determining whether and which biases are indeed present. For example, when not analyzing the articles' content but using the outlets' political orientation \cite{AllSides.com2021} or only the headline \cite{Kong2018}.

In sum, many studies confirm the effectiveness of communicating biases to news consumers. However, prior work suffers from various shortcomings, such as requiring manual analyses, yielding superficial results, or only facilitating the visibility of media bias that \textit{might} be in the data. To address these issues, we propose an approach that automates parts of frame analysis, the established procedure in the social sciences to analyze media bias.

\section{System Description}
\label{sec3}

Given a set of news articles reporting on the same political event, our system seeks to find and visualize groups of articles that similarly frame the event using a three-phase workflow (cf. \cite{Hamborg2019c,Hamborg2020}): article gathering, bias analysis, and bias communication. This section summarizes our previous research concerning article gathering \cite{Hamborg2017a} and bias analysis \cite{Hamborg2019d}. Afterward, we briefly describe the grouping method used in our bias analysis. \Cref{sec4} then introduces our novel visualizations for bias communication. 
For article gathering, we integrate a crawler and extractor for news articles \cite{Hamborg2017a}. Users provide the system with a set of URLs linking to news articles reporting on the same event. The news crawler then extracts the required information from the articles' web pages, i.e., title, lead paragraph, and main text. Alternatively, users can directly provide news articles to the system, e.g., through JSON files.

\begin{figure*}[h]
  \centering
  \includegraphics[width=.9\linewidth]{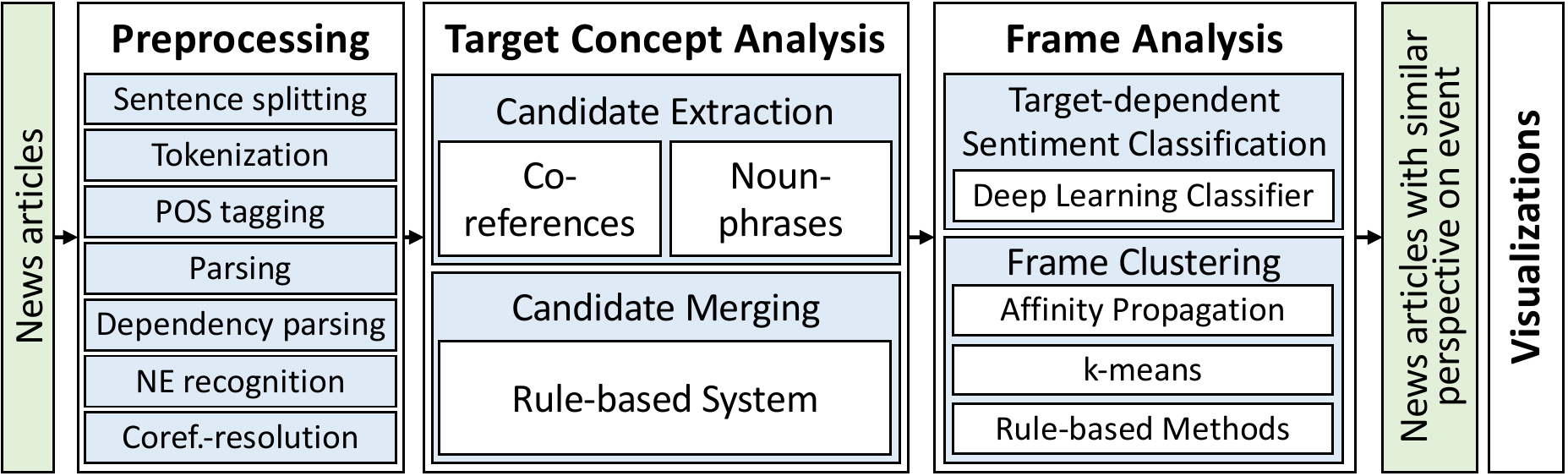}
  \caption{Shown is the three-tasks analysis workflow as it preprocesses news articles reporting on the same event, extracts and resolves phrases referring to persons involved in the event, and groups articles reporting similarly on these persons. Adapted from: \cite{Hamborg2020a}}
  \label{fig:analysisworkflow}
\end{figure*}

Bias analysis performs three tasks as depicted in \Cref{fig:analysisworkflow}. First, we perform state-of-the-art NLP \textit{preprocessing}, including part-of-speech (POS) tagging, dependency parsing, full parsing, named entity recognition (NER), and coreference resolution \cite{Clark2016,Clark2016a}. 
We use CoreNLP with neural models where available and the defaults for the English language \cite{Manning2014}. 
In the following, we describe the subsequent tasks, i.e., target concept analysis and frame analysis.

\textit{Target concept analysis} finds and resolves persons mentioned across the topic's articles, including broadly defined and highly event-specific coreferences as they frequently occur in person-targeting bias forms. Especially in the presence of bias by word choice and labeling, person mentions may be coreferential only in news coverage on a specific event, but otherwise not or even opposing, such as ``freedom fighters'' and ``terrorists'' \cite{Hamborg2019d}. To resolve such mentions, we use the sieve-based method for context-driven cross-document coreference (CDCR) proposed by Hamborg et al. \cite{Hamborg2019d}. 

In sum, our method performs two tasks: candidate extraction and candidate merging. In candidate extraction, we create a base set of coreferential chains from two sources. First, we take the chains from CoreNLP's coreference resolution on the individual news articles (see \textit{preprocessing}). We extend this base set by adding any noun phrase (NP) as singleton coreference chains. The candidate merging task uses six sieves, where each analyzes specific characteristics of two candidates to determine whether they refer to the same semantic concept. For example, the first sieve matches two chains' representative phrases, which represent the chains' core meaning \cite{Manning2014}. The second sieve determines the semantic similarity of all mentions of two chains \cite{Mikolov2013}. The CDCR method achieves a $F1_m=81.7$ compared to $75.3$ achieved by the best baseline, which extracts each noun phrase (NP) and each mention of coreference chains as single candidates chains and clusters the candidate chains in the word2vec space using affinity propagation. Hamborg et al. provide more information on the approach and the baseline \cite{Hamborg2019c}. The output of target concept analysis is the set of persons involved in the news coverage of the event, and for each person, all the person's mention across all news articles.

\textit{Frame identification} determines how news articles portray the persons involved in the event and then finds groups of articles that similarly portray these persons. This task centers around (political) framing \cite{entman1993framing}, where a frame represents a specific perspective on an event. Identifying frames would approximate content analyses, the standard tool used in the social sciences to analyze media bias \cite{Hamborg2019d}. However, doing so would require infeasible effort since researchers in the social sciences typically create frames for a specific research question \cite{entman1993framing,Hamborg2019d}. Our system, however, is meant to analyze media bias on any coverage reporting on policy issues. Thus, we seek to determine a fundamental bias effect resulting from framing: polarity of individual persons, which we identify for each person mention (on the sentence level) and aggregate to article-level. To achieve state-of-the-art performance in target-dependent sentiment classification (TSC) on news articles, we use a RoBERTa-based neural model trained on more than 11k manually labeled sentences sampled from news articles ($F1_m=83.1$) \cite{Hamborg2021b}.

The last step of frame analysis is to determine groups of articles that similarly frame the event, i.e., the persons involved in the event. We propose two methods for grouping. (1) \textit{Grouping-MFA}, a simple, polarity-based method, first determines the single person that occurs most frequently across all articles, called most frequent actor (\textit{MFA}). Then, the method assigns each article to one of three groups, depending on whether the article mentions the MFA mostly positively, ambivalently, or negatively. (2) \textit{Grouping-ALL} considers the polarity of all persons instead of only the MFA. Specifically, the method uses k-means with $k=3$ on a set of vectors where each vector $a$ represents a single news article:
$a = \begin{pmatrix}
s_0 & \cdots& s_{|P|-1}
\end{pmatrix}$
where $P$ is the set of all persons, a person's sentiment polarity $s_i$ in $a$ is
\begin{equation}
    s_i = \sum_{m\in M} w(m)s(m) / m_{\textrm{max},a}
\end{equation}
where $m$ is each mention of all the person's mention in $a$, $w(m)$ is a weight depending on the position of the mention (mentions in the beginning of an article are considered more important \cite{christian2014associated}), $s(m)$ yields the polarity score of $m$ (1 for positive, -1 for negative, 0 else). To consider the individual persons' frequency in an article for clustering, we normalize by $m_{\textrm{max},a}$, which is the number of mentions of the most frequent person in $a$.

In addition to grouping, we calculate each article's relevance concerning the event and concerning the article's group using simple word-embedding scoring.

\section{Visualizations for the Conjoint Experiment}
\label{sec4}
Our visualizations resemble typical online news consumption, i.e., an \textit{overview} enables users to first get a synopsis of news events and articles (\Cref{sec4:overview}) and an \textit{article view} shows an individual news article (\Cref{sec4:articleview}). To measure the effectiveness not only of our visualizations but also their constituents (see our conjoint experiment design \cite{Hainmueller2014} described in \Cref{sec5:methodology}), we design them so that visual features can be altered or exchanged. To more precisely measure the change in bias-awareness concerning only the textual content, we apply changes compared to typical news consumption. For example, the visualizations show the texts of articles (and information about biases in the texts) but no other content, e.g., no photos or outlet names. Further, in our study, the overview shows only a single topic instead of multiple. The visualizations show brief explanations for all visual features.

\subsection{Overview}
\label{sec4:overview}

\begin{figure*}
  \includegraphics[width=.9\textwidth]{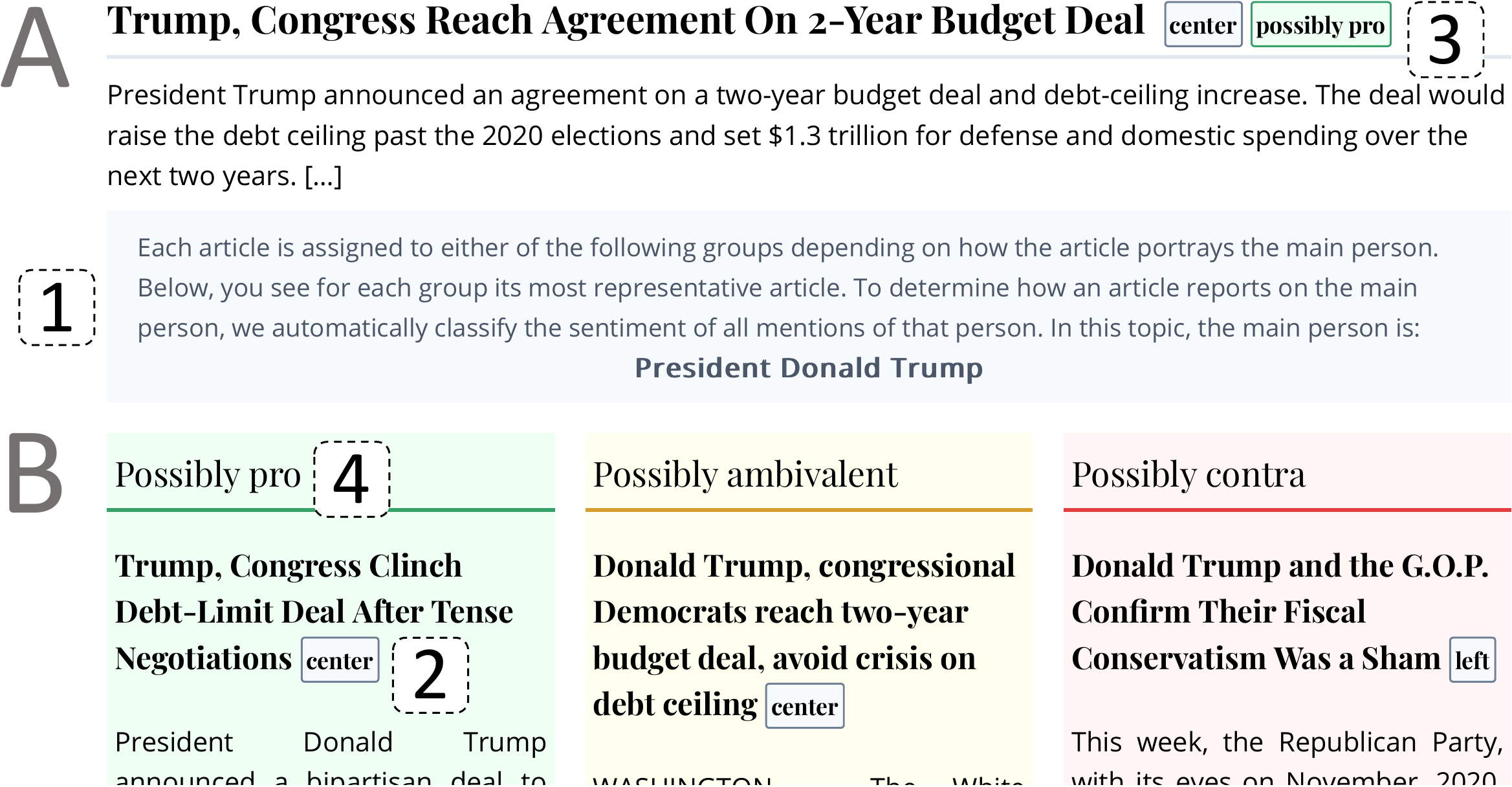}
  \caption{Excerpt of the MFAP overview showing three primary frames present in news coverage on a debt-ceiling event.}
  \label{fig:overview}
\end{figure*}

The overview aims to enable users to get a synopsis of a news event quickly. We devise a modular, bias-sensitive visualization layout, which we use to implement and test specific visualizations. The comparative layout aims to support users in quickly understanding the frames present in coverage of the event. The bias-sensitive layout is vertically divided into three parts (\Cref{fig:overview} shows A and B). The event's \textit{main article} (part A) shows the event's most representative article. The \textit{comparative groups} part (B) shows up to three perspectives present in event coverage by showcasing each perspective's most representative article. To determine the bias-groups, the system uses one of the grouping methods described in \Cref{sec3}, i.e., Grouping-MFA or Grouping-ALL. Finally, a list shows the headlines of \textit{further articles} reporting on the event (bottom, not shown in \Cref{fig:overview}).

In each overview, two types of visual clues conveying bias information can be enabled and altered depending on the conjoint profile (cf. \Cref{sec5:methodology}). First, zero or more headline tags are shown next to each article's headline. They indicate the political orientation of the article's outlet (\textit{PolSides tags}, see ``2'' in \Cref{fig:overview}), the article's overall polarity regarding the MFA due to Grouping-MFA (\textit{MFAP tags}, see ``3''), and the article's group according to its polarity regarding all persons due to Grouping-ALL (\textit{ALLP tags}), respectively. Second, labels and explanations in the visualization are either \textit{generic} or \textit{specific}. The specific variants explain how the grouping was specifically performed (see ``1'') and provide specific group labels (see ``4''). In contrast, all generic variants use the same universal explanation, e.g., only mentioning that our system automatically determined the three perspectives, and use generic coloring and labels, e.g., ``Perspective 1.''

\subsection{Article View}
\label{sec4:articleview}
The article view visualizes a single news article. It thus represents the typically second step in news consumption, i.e., after getting an overview, users subsequently may read individual articles of interest. The article view shows a given article's headline, lead paragraph, and main text. Optionally, the following visual clues to communicate bias information: (1) in-text polarity highlights, (2) polarity context bar, (3) article's bias-groups, and (4) headline tags. These clues are enabled, disabled, or altered depending on the conjoint profile. 

\textit{In-text polarity highlights} aim to enable users to quickly comprehend how the individual sentences of a news article portray the mentioned persons. To achieve this, we visually mark mentions of individual persons within the news article's text. We test the effectiveness of the following modes: \textit{single-color} (visually marking a person mention using a neutral color, i.e., gray, if the respective sentence mentions the person positively or negatively), \textit{two-color} (using green and red colors for positive and negative mentions, respectively), \textit{three-color} (same as two-color and additionally showing neutral polarity as gray), and \textit{disabled} (no highlights are shown). For example, in the sentence ``The Mueller report was tough on Trump,'' the person mention ``Trump'' has a negative polarity and would be highlighted red in the two- and three-colors modes.

\begin{figure}
    \centering
    \includegraphics[width=\linewidth]{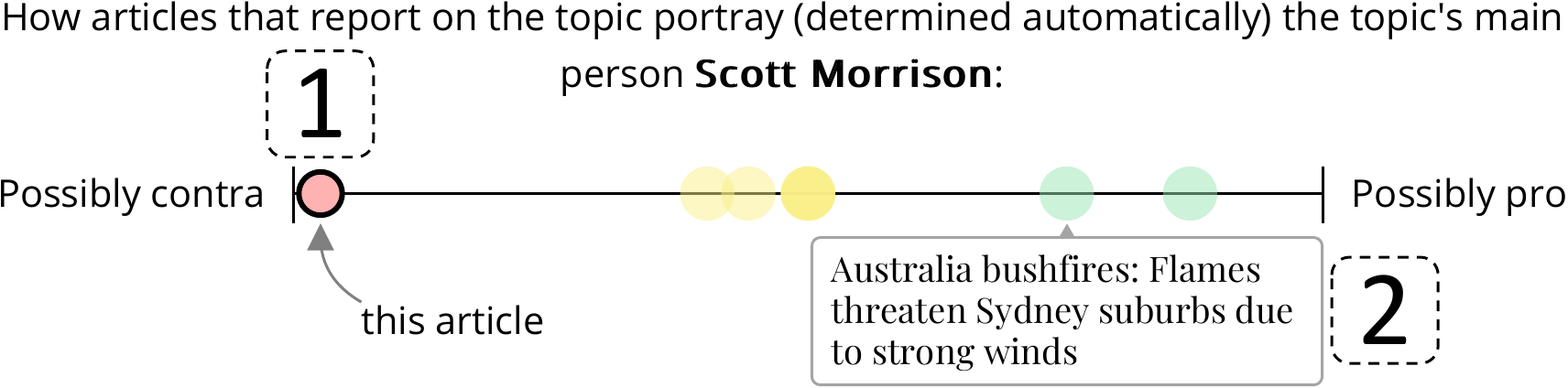}
    \caption{Polarity context bar showing the current and other articles' polarity regarding the MFA and a tooltip showing the headline of a hovered article.}
    \label{fig:polarity-context-bar}
\end{figure}

The \textit{polarity context bar} aims to enable users to quickly contrast how the current article and other articles portray the event's MFA. The 1D scatter plot depicted in \Cref{fig:polarity-context-bar} represents each article as a circle, where the article visualized in the article view is highlighted with a bold circle (see ``1'' in \Cref{fig:polarity-context-bar}). The polarity context bar places each circle depending on its article's overall polarity regarding the MFA. Users can interactively, i.e., by hovering their cursor of circles, view individual articles' headlines (see ``2'').

The \textit{article's bias-groups} part is a set of basic indicators that show the article's bias-group, analogously to the headline tags, i.e., the outlet's political orientation (PolSides), how the article reports on the MFA (MFAP), or all persons (ALLP). In contrast to the headline tags, which are shown besides all headlines, each bias-group indicator is a more prominent component that prominently shows only the bias-group of the article currently shown in the article view. Depending on the conjoint profile (identical to headline tags), individual indicators are shown or disabled. The \textit{headline tags} have identical purpose and function as when shown in the overview (see \Cref{sec4:overview}). 

\section{Study Design}
\label{sec5}

\subsection{Methodology} 
\label{sec5:methodology}
We used a conjoint design \cite{Luce1964}, which is especially suitable for estimating the effect of individual components. Traditional survey experiments are limited to only identifying the ``catch-all effect'' due to confounding of the treatment components \cite{Hainmueller2014}. In contrast, conjoint experiments identify ``component-specific causal effects by randomly manipulating multiple attributes of alternatives simultaneously'' \cite{Hainmueller2014}. In a conjoint design, respondents are asked to rate so-called ``profiles,'' which consist of multiple ``attributes.'' In our study, such attributes are, for example, the overview, which topic it shows (or which article the article view shows), and if or which tags or in-text highlights are shown.

Conjoint experiments rest on three core assumptions: (1) stability and no carry-over effects, (2) no profile-order effects, (3) randomization of the profiles \cite{Hainmueller2014}. In our evaluation, (2) holds by design for all tasks except the forced-choice (see (6) in \Cref{sec5:workflow}). We ensure (3) by randomly choosing the attributes independently of another and for each respondent. To confirm the randomization is successful, we employed a Shapiro-Wilk test \cite{Razali2011} and compared relative frequencies of the demographic dimensions. We found that the data were nearly balanced in each of the three experiments concerning essential demographic dimensions, such as age, sex, education, and political orientation. Thus, also the respondents' demographic data adhere to the conjoint randomization assumption. The political orientation was slightly skewed to the left as is expected with MTurk respondents (cf. \cite{Huff2015}), but still approximates the US distribution \cite{Jones2020}.

To ensure (1), i.e., the absence of carry-over effects from one task set to another, we applied the diagnostics proposed by Hainmueller et al. \cite{Hainmueller2014}. We refer to a task set as all tasks shown to a respondent for a single topic, e.g., in our main study, we show respondents for each topic one overview and subsequently two article views. We then calculate if there are meaningful differences across the task sets by building a sub-group for each task set. We found weak carry-over effects when comparing the individual attributes' effects (using our main overview question across the task sets) and when testing the effect of the task set's order (for all overview questions combined ($\textrm{Est.}=3.28$\%, $p=0.018$), i.e., respondents are on average more bias-aware in the second task set). Further, in our main study, the other attributes' effects differ when sub-grouping for the task set. However, this is not necessarily problematic. A learning effect is expected and desirable in bias communication. Since we randomized the attributes within each task set, we can include the task sets in the analysis and thus control for the effects, regardless of the task set. 

Since in our experiments the three assumptions hold, our design allows for an estimation of the relative influence of each component on the bias-awareness, which is called average marginal component effects (AMCE) \cite{Hainmueller2014}. AMCE ``represents the marginal effect of attribute $l$ [such as a visualization component] averaged over the joint distribution of the remaining attributes'' \cite{Hainmueller2014}.
\comment{and is calculated by 
$\hat{\bar \pi}_l(t_1,t_0,p(\bm{t}))=$
\begin{equation}
\begin{split}
&\sum_{(t,\bm{t}) \in  \mathcal{\widetilde{T}}} \Bigl\{
    \mathbb{E}\bigl[ Y_{ijk} | T_{ijkl} = t_1, T_{ijk[-l]} = t, \bm{T}_{i[-j]k}=\bm{t}\bigr]
\\ &
    -\mathbb{E}\bigl[ Y_{ijk} | T_{ijkl} = t_0,  T_{ijk[-l]} = t, \bm{T}_{i[-j]k}=\bm{t}\bigr]
\Bigr\}
\\ &
\times p\Bigl(T_{ijk[-l]} = t, \bm{T}_{i[-j]k} = \bm{t}  | \bigl(T_{ijk[-l]}, \bm{T}_{i[-j]k}\bigr) \in \mathcal{\widetilde{T}}\Bigr)
\end{split}
\end{equation}
``where $T_{ijk[-l]}$ denotes the vector of $L-1$ treatment components for respondent $i$'s $j$th profile in choice task $k$ without the $l$th component and $\mathcal{\widetilde{T}}$  denotes the intersection of the support of $p(T_{ijk[-l]} = t, \bm{T}_{i[-j]k} = \bm{t}| T_{ijkl}=t_1)$ and $p(T_{ijk[-l]} = t, \bm{T}_{i[-j]k} = \bm{t}| T_{ijkl}=t_0)$. This quantity equals the increase in the population probability that a profile would be chosen if the value of its $l$th component were changed from $t_0$ and $t_1$, averaged over all the possible values of the other components given the joint distribution of the profile attributes $p(\bm{t})$.'' \cite{Hainmueller2014}
}

In our questionnaires, we employ discrete choice (DC) as well as rating questions (cf. \Cref{sec5:workflow}) to measure bias-awareness on a behavioral as well as attitudinal level \cite{Ronen2017}. DC questions are widely used within the conjoint design and found to have high external validity in mimicking real-world behavior \cite{Hainmueller2015}. Additionally, DC questions elicit behavior, i.e., which news article respondents prefer to read or rely on for decision-making \cite{Phillips2002}. In contrast, rating questions capture attitudes and personal viewpoints better \cite{Wijnen2015}.

\subsection{Data}
We selected four news topics with varying degrees of expected biases among the news articles reporting on the topic. To approximate the degree of bias, we used the topics' expected polarization. Specifically, we selected three topics expected to be highly polarizing for US news readers: gun control (Orlando shooting in 2016), debt ceiling (discussions in July 2019), and abortion rights (Tennessee abortion ban in June 2020). To better approximate regular news consumption, where consumers typically are exposed to news coverage on single events, we selected a single event for each of these topics (shown in parentheses). We added a fourth event, which we expected to be only mildly polarizing: Australian bush fires, i.e., a foreign event without direct US involvement.

For each event, we selected ten articles from left-, center, and right-wing online US outlets (political orientation as self-identified by the outlets or from \cite{AllSides.com2021}). We added the abortion topic during our pre-studies due to a strong, negative influence of the debt-ceiling topic on bias-awareness. We could trace this back qualitatively to respondents' critique of the topics being too ``boring'' and ``complicated,'' which also manifested in lower reading times on average. To ensure high quality, we manually retrieved the articles' content. For the second pre-study and our main study, we carefully shortened all articles so that they were of similar length (300--400 words) to address the high reading times and noise in the responses, a key finding of the first pre-study (see \Cref{sec6:prestudies}). We consistently applied the same shortening procedure to preserve the perspectives of the original articles. For example, by maintaining the relative frequency of person mentions and by discarding only redundant sentences that do not contribute to the overall tone. In all experiments, we removed any non-textual content, such as images, to isolate the effects in the change of bias-awareness due to the text content, our text-centric bias analysis, and visualization.

\subsection{Setup and Quality}
\label{sec5:setup}
We conducted our experiments as a series of online studies on Amazon Mechanical Turk (MTurk). To participate in our studies, crowd workers had to be located in the US. To ensure high quality, we further required that participants possess MTurk's ``Masters'' qualification, i.e., have a history of high-quality work. While we compensated respondents always, we discarded data of any respondent who failed to meet all quality criteria, including a minimum study duration, and correctly answering questions checking attention and seriousness \cite{Aust2013}. Depending on the study's duration, participants received compensation that approximated an hourly wage of \$10. 

\subsection{Baselines}
\label{sec5:baselines}
We compare our system and variants with baselines that resemble news aggregators popular among news consumers and an established bias-sensitive news aggregator. Screenshots of their visualizations can be found in the online repository (\Cref{sec1}). \textit{Plain} is an overview variant that resembles popular news aggregators. Using a bias-agnostic design similar to Google News, this baseline shows article headlines and excerpts in a list sorted by the articles' relevance to the event (see relevance calculation described in \Cref{sec3}). \textit{PolSides} is a strong overview variant that represents a bias-sensitive news aggregator \cite{AllSides.com2021}. PolSides yields bias-groups (see ``B'' in \Cref{fig:overview}) by grouping articles depending on their outlets' political orientation (left, center, and right, as self-identified by them or taken from \cite{AllSides.com2021}). Conceptually, PolSides employs the left-right dichotomy, a simple yet often effective means to partition the media into distinctive slants. However, this dichotomy is determined only on the outlet level. It thus may incorrectly classify the biases indeed present in a specific event (see \Cref{sec2:approaches}), e.g., articles shown to be of different slants having indeed similar perspectives (and vice versa). We investigate this issue in our study (see \Cref{sec6}). Because of its similar problem statement, we deem NewsCube \cite{Park2009b} a highly relevant approach. However, we omit it from our study since we suspect the system would achieve poor performance since it was devised using rather quantitative text features. Further, its technique devised for the Korean language cannot be transferred directly to English.

To understand how visualizations, including their layout and explanations, affect bias-awareness compared to what they visualize, i.e., the bias groups resulting from our analysis, we introduce two baseline concepts. First, we include, for most overviews, including the baselines previously mentioned, generic variants (see \Cref{sec4:overview}). Second, we test an overview with generic explanations that randomly assign individual news articles to either of the three bias groups.

For the article view, we test in our conjoint design only the individual visual features described in \Cref{sec4:articleview}, since to our knowledge, there are no easy-to-use visualizations for bias communication in single news articles (see \Cref{sec2}). We test only PolSides tags and MFAP tags and exclude Cluster headline tags in our experiments.

\subsection{Workflow and Questions} 
\label{sec5:workflow}
Our study consists of seven steps. We refer to a \textit{task set} as a sequence of steps associated with one topic, i.e., task set 1 refers to the first topic shown to a respondent, including the overview, the two article views, and respective questionnaires (steps 2--6). The \textit{(1) pre-study questionnaire} asks for demographic and background data \cite{Spinde2020}, such as age, political orientation, news consumption, and attitudes toward the topics we used \cite{Gao2018}. 

Afterward, we show an \textit{(2) overview} as described in \Cref{sec4:overview} and \Cref{sec5:baselines} including instructions shown prior to the overview. The \textit{(3) post-overview questionnaire} then operationalizes the bias-awareness in respondents (see \Cref{sec2:definitions}) by asking about their perception of the diversity and disagreement in viewpoints, if the visualization encouraged them to contrast individual headlines, and how many perspectives of the public discourse were shown, e.g., \textit{Do you think the coverage shown in the previous visualization represents all main viewpoints in the public discourse (independent of whether you agree with them or not) [from Not at all to very much]?  Overall, how did you perceive the articles shown in the previous visualization [very different -- very similar; very opposing -- very agreeing]?}. To match our definition of bias-awareness, we use as our main question (cf. \cite{Park2011}): \textit{When viewing the topic visualization, did you have the desire to compare and contrast articles' [Not at all -- very much]?} 

Afterward, we show an \textit{(4) article view} as described in \Cref{sec4:articleview}. A \textit{(5) post-article questionnaire} operationalizes bias-awareness in respondents on an article-level \cite{Spinde2020}, i.e. \textit{How did you perceive the presented news article? [very unfair -- very fair; very partial -- very impartial; very unacceptable -- very acceptable; very untrustworthy -- very trustworthy; very unpersuasive -- very persuasive; very biased -- very unbiased]}. We also ask whether the article contains political bias and biases against persons mentioned in the article. We repeated steps 2--5 two times since we showed two task sets. After each overview, we showed two articles, i.e., we repeated steps 4 and 5 two times. To measure the effect of seeing an overview before an article, we also introduce a variant where we skip the overview. In such cases, the overview steps (2, 3) are skipped entirely. Afterward, a \textit{(6) discrete choice question} asks respondents to choose between two articles, i.e., which one they consider to be more biased. In a \textit{(7) post-study questionnaire}, respondents give feedback on the study, i.e., what they liked and disliked.

In the two pre-studies, where we tested the study design and usability of the visualizations (see \Cref{sec6:prestudies}), we repeated the same procedure with only one article after each overview and excluding step (5). 

\section{Evaluation}
\label{sec6}

\subsection{Pre-Studies}
\label{sec6:prestudies}
Before our main study, we conducted two pre-studies (E1 and E2) \cite{Hamborg2021q}. E1 consisted of 260 respondents recruited on MTurk (we discarded 3\% discarded from 268 due to the quality criteria described in \Cref{sec5:setup}). E2 consisted of 98 respondents (we discarded 11\% from 110). The pre-studies aimed at testing the study design and usability of the visualizations. We also used the pre-study to find a set of well-performing overviews, including representative baselines. The latter was necessary to satisfy the conjoint assumption \textit{randomization of profiles}, which requires that all profiles have the same set of attributes (each with a randomly selected value for each profile, see \Cref{sec5:methodology}). Across our overviews, the number of attributes differs (cf. \Cref{sec5:baselines}), e.g., Plain has only two attributes (one for each headline tag), our bias-sensitive overview layout (see \Cref{sec4:overview}) has an additional grouping attribute, and ``no overview'' naturally has no attributes (see \Cref{sec5:workflow}). 

In E1, we tested only variants using our bias-sensitive overview layout, where we randomly varied all attributes, i.e., grouping and the two headline tags. We identified (primarily insignificant) trends that indicated well-performing variants. In E2, we then tested the same design as planned for the main study (see \Cref{sec5:workflow}), including the article view and the other baselines (see below).

We also used the pre-studies to improve our design and visualizations. Reasons for partially mixed results in both pre-studies were various usability issues interfering with the effectiveness. For example, in E1, respondents reported they wanted to know how the grouping was performed and by whom. Before conducting E2, we addressed these shortcomings, e.g., by adding explanations (specific and generic) about how our system derives the classifications. After addressing these issues, we found positive, significant effects of our bias-sensitive overviews in the second pre-study, confirming our research design concerning the overview. E2 revealed that headline tags are most effective in improving bias-awareness in the Plain baseline. In contrast, for the bias-sensitive overviews, the bias-awareness remained unchanged or decreased. We suspected that users might feel overwhelmed if many visual clues are present (cognitive load). 

Using the pre-study findings, we defined the following overview variants for the main study. \textit{(1) No overview}. \textit{(2) Plain} as described in \Cref{sec5:baselines}. \textit{(3) PolSides} as described in \Cref{sec5:baselines} with PolSides headline tags enabled to closely resemble the bias-sensitive news aggregator AllSides.com \cite{AllSides.com2021}. \textit{(4) MFA} using the bias-sensitive layout (\Cref{sec4:overview}), Grouping-MFA (\Cref{sec3}), and polarity headline tags enabled, which was the best performing variant of MFA in our pre-studies. \textit{(5) PolSides-generic} being identical to (3) but using generic explanations. \textit{(6) MFA-generic} being identical to (4) but using generic explanations. \textit{(7) Random-generic} using the bias-sensitive layout and random grouping. \textit{(8) ALL-generic} using the bias-sensitive layout, ALL-generic (\Cref{sec3}), and cluster headline tags enabled. Note that we did not test a variant of grouping-ALL with specific explanations.

\subsection{Results}
In our main study, we recruited 174 respondents on MTurk from which we discarded 8\% using our quality measures. In sum, the $n=160$ respondents (age: $[23,77],m=45.5$, 72f/88m/0d, 100\% native speakers, liberal (1)--conservative (10): $m=4.83, sd=2.98$) provided answers to 283 post-overview questionnaires (excluding ``no overview''), 320 discrete choices on article views, and 640 post-article view questionnaires. Our sample size $n<245$ as suggested by Cochran's Formula \cite{Cochran1954,Woolson1986}, but they assume one observation per respondent, whereas we have two observations per respondent. The average study duration was 15min ($sd=6.22$).

Our user study shows that the bias-sensitive overviews strongly and significantly increase respondents' bias-awareness compared to the Plain baseline. The estimate (\textit{Est.}) in \Cref{tab:overviewresults} shows the percentage increase in bias-awareness compared to the attributes' baselines, which is CoreNLP, Plain, and bushfire for the attributes CDCR, Overview, and Topic, respectively. PolSides achieves overall the highest effect when shown with specific explanations ($\textrm{Est.}=21.34$). If shown with generic explanations PolSides has no significant effect ($\textrm{Est.}=8.46$, $p=0.17$). In contrast, our methods for determining bias-groups strongly and significantly increase bias-awareness for specific explanations (MFA: $\textrm{Est.}=17.80$) as well as generic explanations (MFA: $13.35$, ALL: $17.54$). 

\begin{table}[]
\renewcommand{\arraystretch}{1.3}
\renewcommand\cellalign{lc}
\caption{Effects on Bias-Awareness after Overview Exposure}
\label{tab:overviewresults}
\centering
\begin{tabular}{l|l||r|r|r|r}
\hline
\bfseries Attr. & \bfseries Level     & \bfseries Est. & \bfseries SE & \bfseries z  & \bfseries p  \\ \hline \hline
CDCR&TCA                & 1.05 &  2.61 & 0.40  & 0.69       \\ \hline
\multirowcell{6}{Over-\\view} &
Random-gen.              & 6.73  & 5.85 & 1.15  & 0.28      \\ \cline{2-6}
&PolSides            & \textbf{21.34} &  4.84 & 4.40  & ***    \\ \cline{2-6}
&MFA                 & \textbf{17.80}  & 4.90 & 3.63  & ***    \\ \cline{2-6}
&PolSides-gen.   &  8.46 &  6.19 & 1.36  & 0.17    \\ \cline{2-6}
&MFA-gen.        & 13.35 &  5.03 & 2.64  & **     \\ \cline{2-6}
&ALL-gen. & \textbf{17.54} &  5.61 & 3.12  & **    \\ \hline 
\multirow{2}{*}{Topic}&abortion law           & 0.78 &  2.96 & 0.26 & 0.79       \\ \cline{2-6}
&gun control             & 3.16 &  2.94 & 1.07  & 0.25  \\ \hline 
\end{tabular}
\vspace{-0.5cm}
\end{table}

We qualitatively investigated the visualized information of the bias-groups, their articles' content, and respondents' comments. Overall, we found that all grouping methods that achieved significant effects yielded meaningful frames for most topics, especially those polarizing among the political spectrum from left to right. For example, in the gun control topic, the bias-groups achieved by any grouping resemble the frames ``gun control'' and ``gun rights'' with subtle differences between the groups. Grouping-MFA yields two ``gun control'' frames (one being argumentative, the other using factual language) and one ``gun right'' frame (focusing on cruelty and the shooter). In PolSides and grouping-ALL, the ``gun right'' frame focuses on the victims and their right to defend themselves.

However, why is there a loss of effectiveness of PolSides when using generic explanations? Fully elucidating this question would require a larger sample size concerning respondents and topics but we qualitatively and quantitatively identified three potential, partially related causes. \textit{(1) Popularity and intuition:} The left-right dichotomy employed by PolSides is a well-known concept and easily understood by news consumers (none reported they did not understand the concept and 20\% of respondents exposed to PolSides praised that grouping was easy to understand, e.g., ``I liked that it was laid out with left, center, right. It was intuitive.''). In contrast, our grouping techniques are novel and technical, as are their explanations. For example, 30\% of respondents exposed to MFA found the descriptions (slightly) confusing and too ``technical.'' For MFA and ALL, showing generic explanations improved respondents' comprehension, likely because the generic explanations are more conceptual and high-level (10\% for each of MFA-generic and ALL-generic). For PolSides, however, the effect is reversed, potentially indicating a large proportion of the bias-awareness effect is simply due to the well-known dichotomy rather than the visualized bias-groups. 

\textit{(2) Learning effect:} We think that understanding how framing works, i.e. how and through what means news articles can have different perspectives on an event, helps to comprehend the idea of our grouping methods work or at least what they aim to achieve. We hypothesize that while all respondents were aware of the study's focus on media bias (and framing) after the post-overview questionnaire at the end of task set 1, some respondents might not have been before.\footnote{Albeit we also informed respondents about the study's focus on perspectives in the news, at the latest in the post-overview questionnaire they must understand the study's focus to be able to proceed.} This, in turn, would have strongly facilitated comprehending how our grouping methods work or at least what they aim to achieve and thus have a similar effect as described previously for cause (1). While in task set 1, only PolSides employing the well-known left-right dichotomy increased bias-awareness significantly ($\textrm{Est.}=21.55$, $p=.0007$), in the subsequent task set 2, our approaches yielded the strongest, most significant effects (see \Cref{tab:overviewresults-taskset2}). Specifically, MFA achieved the strongest effect ($\textrm{Est.}=28.12$) among the overviews with specific explanations, and the best performing overview with generic explanations (ALL-generic: $26.46$) performed still better than PolSides with specific explanations ($23.54$). 

\begin{table}[]
\renewcommand{\arraystretch}{1.3}
\renewcommand\cellalign{lc}
\caption{Effects after Overview Exposure in Task Set 2}
\label{tab:overviewresults-taskset2}
\centering
\begin{tabular}{l|l||r|r|r|r}
\hline
\bfseries Attr. & \bfseries Level     & \bfseries Est. & \bfseries SE & \bfseries z  & \bfseries p  \\ \hline \hline
CDCR&TCA                & 2.78 &  3.91 & 0.71  & 0.48       \\ \hline
\multirowcell{6}{Over-\\view} &
Random-gen.              & 8.40  & 9.32 & 0.90  & 0.37      \\ \cline{2-6}
&PolSides            & 23.54 &  7.45 & 3.16  & **    \\ \cline{2-6}
&MFA                 & \textbf{28.12}  & 7.03 & 4.00  & ***    \\ \cline{2-6}
&PolSides-gen.   &  12.41 &  8.48 & 1.46  & 0.14    \\ \cline{2-6}
&MFA-gen.        & 21.78 &  7.45 & 2.92  & **     \\ \cline{2-6}
&ALL-gen. & \textbf{26.46} &  8.30 & 3.19  & **    \\ \hline 
\multirow{2}{*}{Topic}&abortion law           & -2.13 &  5.11 & -0.42 & 0.68       \\ \cline{2-6}
&gun control             & 0.85 &  4.80 & 0.18  & 0.86  \\ \hline
\end{tabular}
\vspace{-0.5cm}
\end{table}

\textit{(3) Substantiality of bias-groups:} We qualitatively analyzed the bias-groups yielded by individual grouping methods We found that all grouping methods, including PolSides, determine substantial bias-groups determined in overall for all topics. For some topics, however, the bias-groups determined by MFA and ALL seem to be more substantial compared to bias-groups from PolSides. This is intuitive since PolSides determines bias-groups through the political orientation of the articles' outlets and thus is content-agnostic. In contrast, MFA and ALL analyze in-text features. For example, the debt-ceiling topic employed in our pre-studies highlights this substantiality issue. Our methods yield coherent bias-groups framing the deal negatively, e.g., as political hypocrisy (frame 1), or positively by focusing on (positive) effects for the economy (frame 2) and military (frame 3). In contrast, the bias-groups by PolSides---despite the topic's assumed left-right polarization---resemble rather superficial frames, all framing the deal positively (two groups framing the issue highly similarly, the other focusing on the overall implications of the deal). We suspect that the issue of non-substantial frames is amplified further in news coverage on topics where the left and right wings lack opposing positions.

Overall, the article view does not significantly increase bias-awareness. There are no significant effects for neither component in the article view. Only in-text highlights significantly increase bias-awareness if 3--7 of such highlights are shown and 2-color mode is used ($\textrm{Est.}=2.58$, $p=0.046$). In our qualitative analysis, we identified two potential main causes. \textit{(1) Bias is context-dependent} at least to some degree and thus also depends on relating and contrasting news items \cite{Elejalde2018}. While the bias-sensitive overviews facilitate contrasting news articles, the article view shows only a single news article. Further, the overall tone of an article, e.g., as visualized in the overview, might be more important for bias-perception than individual features (cf. \cite{Feick2021}). \textit{(2) User experience (UX) issues:} When qualitatively reviewing respondents' feedback, we identified various UX issues, especially when too many visual clues were shown, e.g., 3-colors mode or many in-text highlights may have caused mental overload \cite{Rogowitz1998}, and 30\% of the respondents that had more than seven highlights reported that they felt ``overwhelmed.'' We also identified issues of other components, e.g., the polarity context bar did not increase bias-awareness albeit conceptually facilitating the comparison of news articles. Often, many articles were placed as circles on the same spot (due to the grouping-MFA). Further, it prominently visualizes only the articles' MFA polarity, but users would need to see actual content rather than such a derived characteristic.

We identified two potential causes concerning why highlights were only effective when their frequency was in a specific range. An article with less than three highlights might not contain enough sentiment to be perceived as biased. On the other hand, showing more than seven highlights could lead to a ``mental overload'' could be reached \cite{Rogowitz1998}. For example, 30 \% of respondents reported feeling ``overwhelmed'' by the highlights and thus could not derive a consistent conclusion on whether the article contained bias. 

Showing an overview before the individual news articles had inconclusive effects. We found a significant, mild effect caused by only the MFA overview ($\textrm{Est.}=4.64$, $p=0.003$); other overviews had no significant effects. 

\subsection{Limitations and Future Work}
In our view, the main limitations of our experiments and results are their \textit{representativeness and generalizability}, mainly due to three partially related factors. \textit{(1) Study design,} e.g., respondents had to view given events and articles rather than deciding what to read. An interactive design might more closely resemble real-world news consumption and address further issues we faced in our experiments, such as study fatigue. While our study's duration is well below where one would expect study fatigue \cite{Schatz2012}, users on MTurk typically work on many online tasks. Rather than querying respondents for the subjective concept of bias-awareness, a long-term usage study could directly measure the effects of our approach on news consumption. For example, if respondents will read more articles portraying events from different perspectives \cite{Park2009b}.

\textit{(2) Respondent sample:} While our sample approximately resembles the US distribution concerning a dimension important in this study, i.e., political affiliation (cf. \cite{PewResearchCenter2020,Perloff2015}), the sample contains selection biases, e.g., since we recruited respondents only on one platform and from only the US. Thus, we cannot conclude findings for other news consumers of countries with systematically different political or media landscapes. For example, while the two-party system may lead to more polarizing news coverage in the US, countries with multi-party systems typically often have more diversified media landscapes \cite{Yang2016}. Further, we seek to increase the sample size since we found inconclusive or insignificant effects for respondents' demographics, their attitudes toward the topics, and in task sets or other sub-groups.

Our \textit{(3) event and article sample} yields similar limitations as described for (2) due to its small size and systematic creation. We suggest increasing the number of events and articles per event and use a random sample. Lastly, our study did not relate bias-awareness to the articles' content but only to our approach, the respondents, and a topic's expected degree of polarization. To measure the effect of content and biases therein, a future study could relate bias-awareness to a ground-truth created using manual frame analysis.

Besides the previously mentioned limitations and future improvements concerning our study, we plan to address the following issues concerning our approach and the usability of our visualizations. We think that our article view's inconclusive effects are partly due to non-optimal UX, e.g., the view may visualize too few or too many in-text highlights (5\% article views did not contain any in-text highlights) or ineffective visual clues, such as the polarity context bar. Another reason for the inconclusive effectiveness might be because bias-awareness is a comparative concept, and the article view shows only a single article. We think that showing representative summaries of each bias group will increase the overview's effectiveness compared to showing the headlines of representative articles. Headlines may contain journalistic hooks and, if at all, are representative for their article but not the group. While our approach yields high effectiveness overall, we also found that our analysis is sub-optimal in topics that are not person-oriented, e.g., in the bushfire topic, where news coverage focused on the consequences for society, economy, and nature. We seek to extend our analysis to other semantic concepts and investigate topic-independent frames and derivations (cf. \cite{Hamborg2019c,Kwak2020,Card2015}).

\section{Conclusion}
We present the first system to automatically identify and then communicate person-targeting forms of bias in news articles reporting on policy issues. Earlier, researchers could reliably identify these biases only using content analyses, which---despite their effectiveness in capturing also subtle biases---could only be conducted for few topics in the past due to their high cost, manual effort, and required expertise. In a large-scale user study, we employ a conjoint design to measure the effectiveness of visualizations and individual components. We find that our overviews strongly and significantly increase bias-awareness in respondents. In particular, and in contrast to prior work, our bias-identification method seems to reveal biases that emerge from the content of news coverage and individual articles. In practical terms, the study results suggest that the biases found and communicated by our method are indeed \textit{present} in the news articles, whereas the reviewed prior work rather \textit{facilitates} the detection of biases, e.g., by distinguishing between left- and right-wing outlets.

\section*{Acknowledgment}
This work was partially funded by the WIN program of the Heidelberg Academy of Sciences and Humanities, financed by the Ministry of Science, Research and the Arts of the State of Baden-Wurttemberg, Germany. We thank the anonymous reviewers for their valuable comments.

%\section*{References}
\bibliographystyle{IEEEtran}
\bibliography{refsfrozen}

\end{document}